\documentclass[useAMS,usenatbib]{mn2e}
\usepackage[]{natbib,amsmath,amssymb,times}
\bibpunct{(}{)}{;}{a}{}{,}
\usepackage{graphicx}
\usepackage{epsfig}
\usepackage{latexsym}
\usepackage{ulem}

\def\reff@jnl#1{{\rm#1\/}}
\def\aj{\reff@jnl{AJ}}                  
\def\araa{\reff@jnl{ARA\&A}}            
\def\apj{\reff@jnl{ApJ}}                
\def\apjl{\reff@jnl{ApJ}}               
\def\apjs{\reff@jnl{ApJS}}              
\def\apss{\reff@jnl{Ap\&SS}}            
\def\aap{\reff@jnl{A\&A}}               
\def\aapr{\reff@jnl{A\&A~Rev.}}         
\def\aaps{\reff@jnl{A\&AS}}             
\def\mnras{\reff@jnl{MNRAS}}            
\def\prd{\reff@jnl{Phys.Rev.D}}         
\def\prl{\reff@jnl{Phys.Rev.Lett}}      
\def\pasp{\reff@jnl{PASP}}              
\def\pasj{\reff@jnl{PASJ}}              
\def\nat{\reff@jnl{Nature}}             
\def\physrep{\reff@jnl{Phys. Rep.}}     

\newcommand{\bea}{\begin{eqnarray}}
\newcommand{\eea}{\end{eqnarray}}
\newcommand{\be}{\begin{equation}}
\newcommand{\ee}{\end{equation}}
\newcommand{\rund}[1]{\left(#1\right)}
\newcommand{\vc}[1]{\mbox{\boldmath $#1$}}
\renewcommand{\d}{{\rm d}}
\newcommand{\dc}{\partial}
\newcommand{\eck}[1]{\left[ #1 \right]}

\title[Likelihood of shear and flexion]
{Measurement of halo properties with weak lensing shear and flexion}
\author%
[Er et al.]%
{Xinzhong Er$^{1}$\thanks{E-mail:xer@nao.cas.cn},
  Ismael Tereno$^{2}$,
  and
  Shude Mao$^{1,3}$
\\
  $^1$National Astronomical Observatories, Chinese Academy of Sciences,
  Beijing 100012, China\\
  $^2$Centro de Astronomia e Astrofisica da Universidade de Lisboa,
  Tapada da Ajuda, 1349-018, Lisboa, Portugal \\
  $^3$Jodrell Bank Centre for Astrophysics, University of Manchester,
  Alan Turing Building, Manchester M13 9PL, UK
}
\date{Accepted 2011 December 16; received 2011 December 15; in original form 2011 August 13}

\pubyear{2011}

\begin{document}

\maketitle

\begin{abstract}
  We constrain properties of cluster haloes by performing likelihood
  analysis using lensing shear and flexion data. We test our analysis
  using two mock cluster haloes: an isothermal ellipsoid (SIE) model
  and a more realistic elliptical Navarro-Frenk-White (eNFW)
  model. For both haloes, we find that flexion is more sensitive to
  the halo ellipticity than shear.  The introduction of flexion
  information significantly improves the constraints on halo
  ellipticity, orientation and mass. We also point out that there is a
  degeneracy between the mass and the ellipticity of SIE models in the
  lensing signal.
\end{abstract}%
\section{Introduction}

The properties of galaxy and cluster haloes are of great interest in
cosmology, and can be powerful tests of the cosmological paradigm and
the nature of dark matter. Two important parameters that describe a
dark matter halo are its mass and shape, which are related to many
physical processes, such as the growth and merging history
\citep{1993MNRAS.264..201K,2005Natur.435..629S}.

Models of dark matter haloes beyond the spherical approximation are
favored by many numerical simulations \citep{2002ApJ...574..538J,
  2004IAUS..220..421S,2004ApJ...611L..73K, 2006MNRAS.367.1781A} and
observations \citep{2000A&A...364..377R, 2004ApJ...601..599L,
  2005ApJ...625..108D, 2006ApJ...645..170S,
  2010ApJ...718..762W}. Furthermore, numerical simulations with
different assumptions predict different properties of dark matter
haloes \citep[e.g.][]{2002sgdh.conf..109B,
  2005ApJ...627..647B,2007MNRAS.380...93W}.  Current models based on
N-body simulations, semi-analytic models or hydrodynamic simulations
can predict several halo properties, but several
ingredients of these models remain uncertain. A precise understanding of
halo properties such as mass and ellipticity is important to confirm and
improve the existing models of galaxy formation and probe the
physical nature of dark matter.

Gravitational lensing is a powerful tool to study mass distributions,
independent of the nature or dynamical state of the matter
\citep[see][for reviews]{1999ARA&A..37..127M,2006glsw.book..269S,
  2008PhR...462...67M}. Galaxy-Galaxy Lensing (GGL) is concerned with
the mass associated with galaxies and dark matter haloes in which galaxies
reside \citep{1984ApJ...281L..59T, 1996ApJ...466..623B,
  1998ApJ...503..531H, 2004ApJ...606...67H, 2004AJ....127.2544S,
  2006MNRAS.368..715M}. The distortion caused by a single galaxy
cannot be detected, but the statistics of many foreground-background
pairs yield a detectable signal for a population of galaxies.
\citet{1996ApJ...466..623B} discovered a significant GGL shear signal.
\citet{1997ApJ...474...25S} developed a maximum likelihood analysis
that can constrain the halo properties of the lens galaxy populations
through GGL, allowing to estimate the mean velocity dispersion and the
characteristic scale for a non-singular isothermal sphere halo model.

Flexion as the gradient of the projected mass density, is sensitive to
the small-scale variations of mass distributions
\citep{2002ApJ...564...65G,2005ApJ...619..741G,2006MNRAS.365..414B}.
Different techniques have been developed to measure flexion
\citep[see][for examples]{2006ApJ...645...17I, 2007ApJ...660..995O,
  2008A&A...485..363S,2011MNRAS.416.1616F}. Recently,
\citet{2011MNRAS.412.2665V} applied the shapelets technique on the
COSMOS survey, and \citet{2011ApJ...736...43C} introduced a new method
(so-called analytic image model) to analyse lensing flexion images. It
has been noted that flexion can contribute to cosmology in several
aspects, such as exploring the mass distribution of dark matter haloes
of galaxies and clusters, especially substructures
\citep{2009MNRAS.395.1438L, 2010MNRAS.409..389B, 2010arXiv1008.3088E}.
\citet{2011MNRAS.412.1023H} also propose to reduce the distance
measurement errors of standard candles using lensing shear and flexion
maps.

In \citet{2009MNRAS.400.1132H, 2011A&A...528A..52E,
 2011MNRAS.417.2197E}, the ellipticity of a galaxy halo has been
studied with flexion. It was found that the constraints from flexion
are tighter than those from shear. In \citet{2005ApJ...619..741G},
Galaxy-Galaxy lensing Flexion (GGF) has been studied using the
distribution function of orientations between the line connecting
foreground and background pairs and the flexion of the background
galaxies. A GGF signal has been detected by
\citet{2007ApJ...666...51L} using images taken by HST ACS in the
cluster Abell 1689. Moreover, combining shear and flexion information
provides tighter constraints on halo properties by studying mass
distribution on different scales
\citep{2010arXiv1008.3088E,2010MNRAS.404..858S,2011MNRAS.412.1023H}.

In this paper, we combine shear and flexion data to constrain the
properties of dark matter haloes. The tangential shear is mainly
sensitive to the mass of the halo, whereas flexion is sensitive to the
halo ellipticity.  Moreover the usable number density of flexion data
is relatively low, since the flexion signal drops faster than the
shear signal with the angular distance to the centre of the halo. It
is thus not sufficient to constrain the halo properties with shear or
flexion alone. Therefore, we propose to take advantage of both shear
and flexion in our analysis. In particular, we use the angular
positions, tangential shear and flexion of galaxies in our likelihood
functions. A singular isothermal ellipsoid model and an elliptical
NFW model for a galaxy or cluster halo with 3 parameters (mass,
ellipticity and orientation) are adopted in this paper. In Sec. 2, we
recall the basic lensing equations. Our likelihood function is
introduced in Sec. 3. We perform numerical tests of our method and results in
Sec. 4 and present our conclusions in Sec. 5.  Throughout this paper,
we adopt a $\Lambda$CDM model with $\Omega_{\Lambda}=0.75$,
$\Omega_{\rm m}=0.25$, and a Hubble constant $H_0 = 73$
km\,s$^{-1}$\,Mpc$^{-1}$.

\section{\label{Sc:2}Lensing Basics}

The formalism described here can be found in
\citet{2008A&A...485..363S, 2011A&A...528A..52E}.  The weak lensing
shear and flexion are conveniently described using a complex
formalism. We adopt the thin lens approximation, assuming that the
lensing mass distribution is projected onto a single lens plane. The
dimensionless projected mass density can be written as $\kappa(\vc
\theta)= \Sigma(\vc\theta)/ \Sigma_{\rm cr}$, where $\vc\theta$ is the vector of
(angular) position coordinates, $\Sigma (\vc\theta)$ is the projected
mass density and $\Sigma_{\rm cr}$ is the critical density, given by
\be
\Sigma_{\rm cr}= {c^2 \over 4\pi G} {D_{\rm s}(\infty)\over
D_{\rm d} D_{\rm d,\infty}},
\ee
for a fiducial source located at a redshift $z \to \infty$. Here
$D_{\rm s}(\infty)$, $D_{\rm d}$ and $D_{\rm d,\infty}$ are the angular
diameter distances between the observer and the source, the observer and the
lens and between the lens and the source, respectively.

The first order image distortion induced by gravitational lensing is
the shear $\gamma$, which transforms a circular source into an
elliptical one. The second order effect, called flexion, is described
by two parameters: the spin-1 flexion, which is the complex derivative
of $\kappa$
\be
{\cal F}= \nabla_{\rm c} \kappa= {\dc \kappa \over \dc\theta_1}
+ {\rm i}{\dc\kappa \over \dc\theta_2},
\ee
and the spin-3 flexion, which is the complex derivative of $\gamma$
\be
{\cal G}= \nabla_{\rm c} \gamma.
\ee

For a source at redshift $z_{\rm s}$ and a lens at redshift $z_{\rm d}$, a
`cosmological weight' function must be introduced:
\be
Z(z_{\rm s})\equiv \eck{D_{\rm d,\infty} \over D_{\infty}}^{-1}{ D_{\rm ds} \over
  D_{\rm s}} \,H(z_{\rm s}-z_{\rm d}),
\label{cosmoweight}
\ee
where $D_{\rm ds}$ and $D_{\rm s}$ are the angular diameter distances
between the lens and the source, and the observer and the source.
$H(z_{\rm s}-z_{\rm d})$ is the Heaviside step function to ensure that the
source redshift is higher than the lens redshift. The first and
second-order lensing effects scale with the source redshift as
\bea
\kappa(z_{\rm s}) = Z(z_{\rm s}) \kappa, &\,& \gamma(z_{\rm s}) =
Z(z_{\rm s}) \gamma, \nonumber\\
{\cal F}(z_{\rm s}) = Z(z_{\rm s}) {\cal F}, &\,& {\cal G}(z_{\rm s})
= Z(z_{\rm s}) {\cal G}.
\label{kz}
\eea

In GGL we express the shear with respect to a foreground
halo. This defines the tangential shear
\be
\gamma_{\rm t}=-\gamma_1\cos2\psi - \gamma_2\sin2\psi,
\label{tshear}
\ee
where $\gamma_1$, $\gamma_2$ are respectively the real and the
imaginary components of shear, and $\psi$ is the polar angle with
respect to the vector connecting the background and foreground
galaxies. ${\cal F}_1$ (${\cal G}_1$) and ${\cal F}_2$ (${\cal
  G}_2$) are the real and imaginary components of the spin-1 (spin-3)
flexion.
%
%

In this paper, $\gamma_{\rm t}$, ${\cal F}_1$, ${\cal F}_2$, ${\cal
  G}_1$ and ${\cal G}_2$ represent the values calculated from our
model, while $e$, ${\cal F}^{\rm obs}_{1}$, ${\cal F}^{\rm obs}_{2}$,
${\cal G}^{\rm obs}_{1}$ and ${\cal G}^{\rm obs}_{2}$ are the
observables, which will be introduced in the next section.

\section{Methodology}
\label{sect:likelihood}
We apply the Bayesian framework to study galaxy-galaxy lensing simulated
maps, in order to estimate physical parameters of a foreground dark
matter halo. The maps contain tangential shear and two
flexion components at the positions of background galaxies. The values of
these observables reflect the intrinsic shapes of the background
images, slightly modified by lensing due to the gravitational
potential of a foreground halo.

We assume the shear and flexion components can be measured with
unbiased estimators that are linear combinations of the lensing and
intrinsic contributions to an image shape. The ellipticity of
background galaxies $e$ is such an estimator. Analogously, we assume
that the estimators of flexion, ${\cal F}^{\rm obs}_{1}$,
${\cal F}^{\rm obs}_{2}$, ${\cal G}^{\rm obs}_{1}$ and ${\cal G}^{\rm obs}_{2}$
are derived from higher-order brightness moments of the
background images. We further assume that the noise of such estimators
is uncorrelated due to the distributions of intrinsic
shapes, and neglect other contributions, such as sample variance.


The intrinsic ellipticity distribution can be described by a Gaussian
probability density distribution with zero mean and standard deviation
$\sigma_e\approx0.3$ \citep{1996ApJ...466..623B}.
For flexion, we use the preliminary studies of intrinsic flexion by
\citet{2005ApJ...619..741G} and \citet{2007ApJ...660.1003G}, where
scatter of $\sigma_{\cal F}=0.03$ per arcsecond and
$\sigma_{\cal G}=0.04$ per arcsecond were found for spin-1 and spin-3 flexion, respectively.
In this paper, we also assume that the
distribution of each intrinsic flexion noise component
$n_{Fj}={\cal F}^{\rm obs}_{j} - {\cal F}_j$,
$n_{Gj}={\cal G}^{\rm obs}_{j} - {\cal G}_j$
($j=1,2$) is a Gaussian with zero mean and $\sigma_{\cal F}=0.03/''$,
$\sigma_{\cal G}=0.04/''$
\bea
P(n_{Fj}) = {1 \over \sqrt{2\pi} \sigma_{\cal F}}
{\rm exp}\eck{-\frac{n_{Fj}^2}{2\sigma_{\cal F}^2}}\,,\\
P(n_{Gj}) = {1 \over \sqrt{2\pi} \sigma_{\cal G}}
{\rm exp}\eck{-\frac{n_{Gj}^2}{2\sigma_{\cal G}^2}}\,.
\label{frnoisedis}
\eea

Constraints on the foreground haloes parameters $\vc p$, which will be
introduced in the next section, are inferred by evaluating their
likelihood functions. The likelihood of a model is the conditional probability
of the data given the model. Following the previous discussion, we
define three likelihood functions:
\bea
L_e (\gamma_i({\vc p}))&=& {1\over \sqrt{2\pi}\,\sigma_e} {\rm exp}\eck{-(e_i -\gamma_i)^2
\over 2\sigma_e^2},\\
L_{F}({\cal F}_{i}({\vc p})) &=& {1\over \sqrt{2\pi}\,\sigma_{\cal F}}
{\rm exp}\eck{-({\cal F}^{\rm obs}_{i}-{\cal F}_{i})^2\over 2\sigma_{\cal F}^2},\\
L_{G}({\cal G}_{i}({\vc p})) &=& {1\over \sqrt{2\pi}\,\sigma_{\cal G}}
{\rm exp}\eck{-({\cal G}^{\rm obs}_{i}-{\cal G}_{i})^2\over 2\sigma_{\cal G}^2}.
\eea
The subscript $i$ refers to the $i^{\rm th}$ background galaxy.
Notice that not all images have a measurable shape and
usually flexion is measurable only for a subset of all the images. For this
reason, a selection must be applied to the data. The details of the mock data
produced are given in the next section.

The galaxy shapes are assumed to be independent, i.e., in this study
we assume uncorrelated spatial noise, and no systematic intrinsic
shape correlations or spurious correlations from PSF residuals. Each
of the three likelihoods can thus be multiplied over all galaxy pairs
with measurable background galaxy shape information. Furthermore, if we
assume that the measurements of tangential shear and
flexion are independent at each galaxy position, the likelihood
can be written as
\be
L= \eck{\prod_i (L_e(\theta_i))} \eck{\prod_i(L_{F}(\theta_i))}
\eck{\prod_i(L_{G}(\theta_i))}.
\label{likeliall}
\ee
Recently \citet{2011arXiv1107.3920V} pointed out a
correlation between shear and flexion noises. Although the correlation
can be reduced, a precise covariance treatment of shear and
flexion will require more detailed empirical knowledge of the flexion noise
and is beyond the scope of this paper.

\section{Analysis and Results}
\label{sect:test}
\begin{figure}
\centerline{\scalebox{1.0}
  {\includegraphics[width=7.5cm,height=6.0cm]{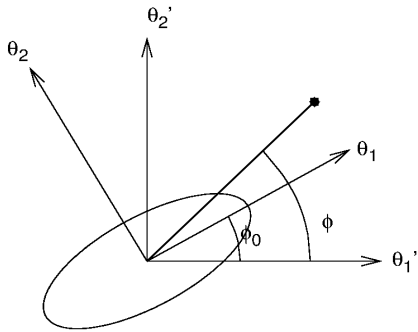}}}
  \caption{The coordinate systems and relative angles used. The star
    represents the location of one background galaxy.  }
  \label{fig:coors}
\end{figure}
In this section, we first use the analytic singular isothermal ellipsoid
(SIE) model to illustrate the results before we study the more realistic
elliptical Navarro-Frenk-White (eNFW) profile.

\subsection{SIE model}
We adopt an SIE model for
the foreground lens. The SIE model includes
3 parameters: the Einstein radius $\theta_{\rm E}$, which defines the
scale of the lens and is related to the mass or velocity dispersion
(see Eq.~\ref{thetae}); the halo ellipticity, defined by
$\epsilon=(\theta_a-\theta_b)/(\theta_a+\theta_b)$ (or equivalently by
the axial ratio $f=\theta_b/\theta_a)$, where $\theta_a,\theta_b$ are
the major and minor axes; and the halo orientation $\phi_0$.

The lensing properties of the halo, such as shear and flexion, are
calculated at the background galaxies positions
$(\theta_1',\theta_2')$. The image reference frame relates to the halo
reference frame through the rotation
$\theta_1=\theta_1'\cos\phi_0+\theta_2'\sin\phi_0$,
$\theta_2=-\theta_1'\sin\phi_0+\theta_2'\cos\phi_0\,$, where $\phi_0$
is the halo orientation (see Fig.~\ref{fig:coors} for an
illustration).

The dimensionless surface
mass density produced by an SIE halo at the location $(\theta_1',\theta_2')$
of a background galaxy is given by
\be
\kappa (\theta_1',\theta_2'|\theta_{\rm E},\epsilon,\phi_0)=
{\theta_{\rm E} \over \rho},
\label{kappa}
\ee
with $\rho(\theta_1,\theta_2)$ defined by
\be
\rho=\sqrt{\theta_1^2 f^2+\theta_2^2}.
\label{rho}
\ee

The halo
produces the following shear and flexion fields for a source at $z_s=\infty$:
\bea
{\gamma (\theta_1',\theta_2') \over\theta_{\rm E}} &=& -
{\theta_1^2-\theta_2^2 \over \rho\theta^2}
-{\rm i}\; {2\theta_1\theta_2\over \rho\theta^2};
\label{isodensityg}\\
{{\cal F} (\theta_1',\theta_2')\over \theta_{\rm E}} &=&
-{\theta_1 f^2 \over \rho^3} -{\rm i}\; {\theta_2 \over \rho^3},
\label{isodensityf}\\
{{\cal G } (\theta_1',\theta_2')\over \theta_{\rm E}}
&=& -\rund{{2\theta_1^3-6\theta_1\theta_2^2
\over \theta^4\rho} + {f^2\theta_1^3 - f^2\theta_1\theta_2^2 -
2 \theta_1\theta_2^2 \over \theta^2\rho^3}} \nonumber\\
&&-{\rm i} \rund{{6\theta_1^2\theta_2 - 2\theta_2^3 \over
\theta^4\rho} + {\theta_1^2\theta_2 - \theta_2^3 + 2 f^2 \theta_1^2\theta_2
\over \theta^2\rho^3}},
\label{isodensityf3}
\eea
where $\theta=\sqrt{\theta^2_1+\theta_2^2}$ and $f$ is the axial ratio.
The shear and flexion will
be scaled to different redshifts for other values of $z_s$, according
to Eq.~(\ref{kz}).

\begin{figure*}
\centerline{\scalebox{1.0}
  {\includegraphics[width=7.5cm,height=7.cm]{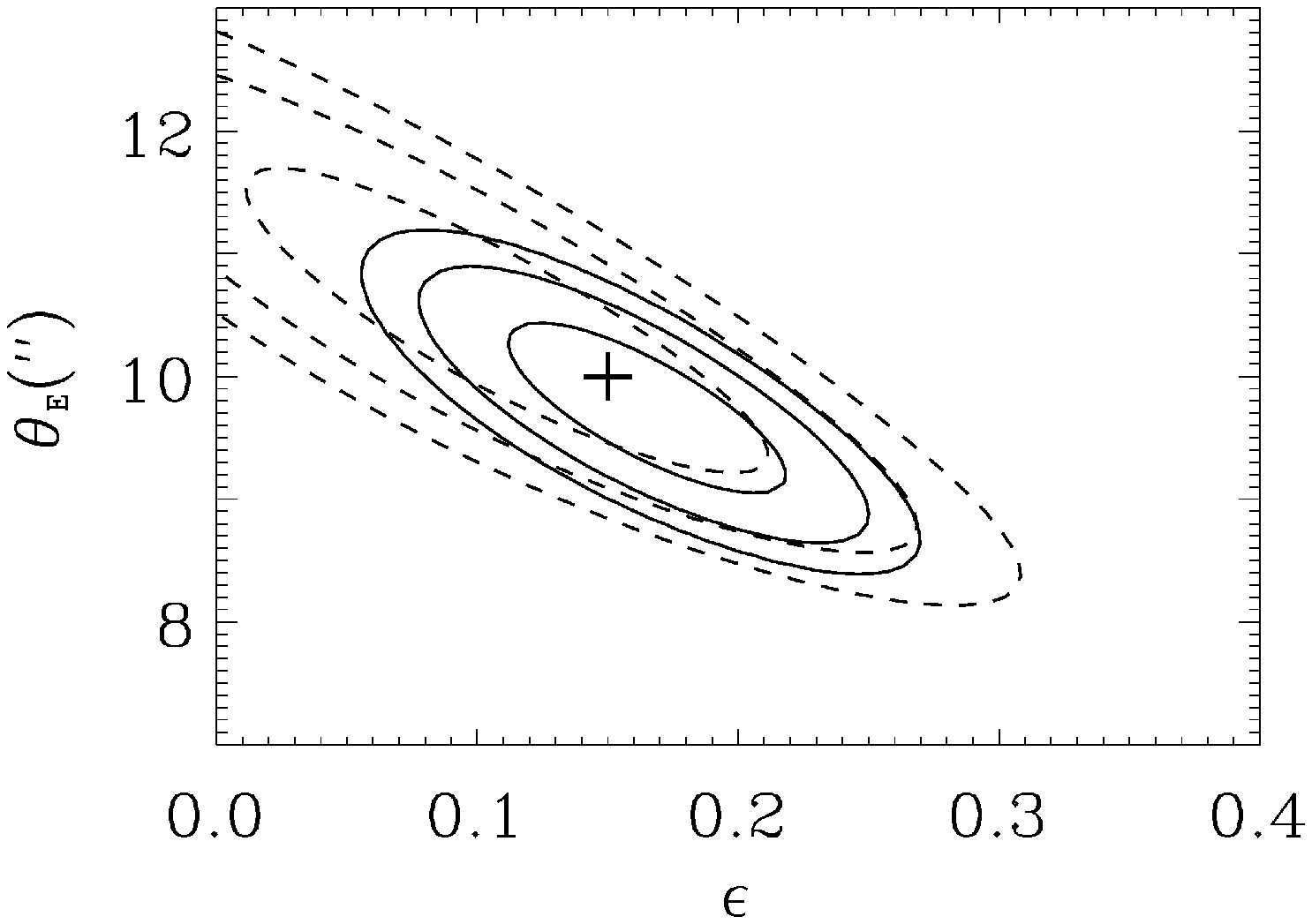}
    \includegraphics[width=7.5cm,height=7.cm]{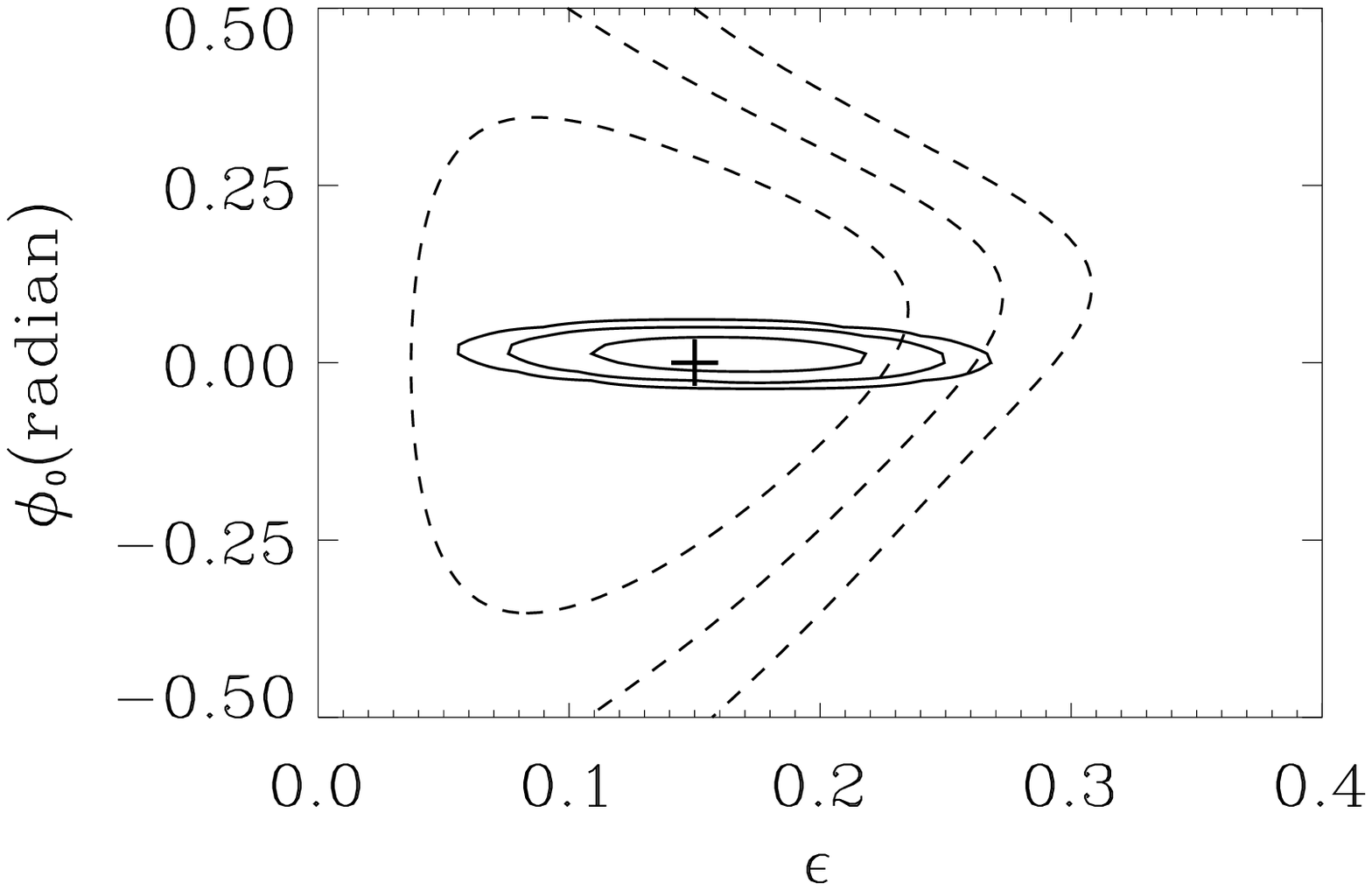}}}
  \caption{Marginalized SIE $68\%$, $95\%$ and $99\%$ credible regions using
    shear data only (dashed) and combined shear + flexion data (solid).
    Left panel: $(\theta_{\rm E}$-$\epsilon$) plane. Right panel: $(\phi_0$-$\epsilon)$ plane. The cross shows the fiducial input model. }
  \label{fig:shearonly}
\end{figure*}

\subsection{Simulated fields}
We place a halo, with fiducial parameter values $\theta_{\rm E} =10''$,
$\epsilon=0.15$, $\phi_0=0$ and redshift $z_{\rm d}=0.6$, at the center of a
$1.5'\times 1.5'$ field.
The Einstein radius $\theta_{\rm E}$ for a
singular isothermal spherical halo with velocity dispersion $\sigma_v$
is given by
\be
\theta_{\rm E}=4\pi\rund{\sigma_v \over c}^2 {D_{\rm ds}\over D_{\rm s}}.
\label{thetae}
\ee
Thus for the case of $\theta_{\rm E} =10''$ and $z_{\rm s}=1.45$
(implying $D_{\rm ds}/D_{\rm s}=0.5$),
the velocity dispersion is about $840$ km/s, corresponding
to a large group or a cluster.

We then place 80 background galaxies in the field at random positions.
A redshift is assigned to each galaxy according to a Gamma
distribution with $z_0=1/3$,
\be
p(z)={z^2 \over 2z_0^3} {\rm exp}\rund{-{z\over z_0}},
\label{redshift}
\ee
which peaks at $z=2/3$ and has a mean redshift of $\langle z\rangle
=3z_0=1$. The source density corresponds to space-based observing
conditions. It is compatible with the density of the weak lensing
source galaxies observed in the COSMOS field
\citep{2010A&A...516A..63S}.

The lensing shear and flexion components are computed for each galaxy
according to the formula given in the previous section and using
the cosmological weight function
defined in Eq.~(\ref{cosmoweight}).

We add noise to the shear and flexion signals. The shear noise is
generated from a Gaussian distribution with $\sigma_e=0.2$ per shear
component, and the flexion noise is generated from a Gaussian
distribution with $\sigma_{\cal F}=0.03/''$ and $\sigma_{\cal
  G}=0.04/''$ per flexion component. Galaxies with shear absolute
value larger than 0.9 are discarded from the analysis. This
corresponds to about a few percent of the sample.  We further discard,
from the flexion analysis: 1) galaxies with flexion absolute value
larger than $0.5/''$, since correct flexion estimates cannot be
obtained in very strongly distorted images (see Schneider \& Er
2008). 2) galaxies with small flexions ($|{\cal F}|<0.005/''$), which
are exceedingly difficult to measure due to their low signal-to-noise
ratios. 3) galaxies located at high redshift ($z>1.0$), which are too
faint or too small to allow for a reliable flexion measurement
\citep{2008ApJ...680....1O}. 4) galaxies located at low redshift
($z<0.61$) are also discarded since they are either below the lens
redshift or too close to the lens to be efficiently lensed. In the
end, we discard about $70\%$ of the flexion data in our analysis.

\subsection{SIE results}

We perform a likelihood analysis to assess how well
the fiducial model can be recovered given the noisy shear and the
two noisy flexion components in the simulated data.

The theoretical predictions are
calculated using Eqs.~(\ref{isodensityg})-(\ref{isodensityf3}) and the
three parameters  $\theta_{\rm E}$, $\epsilon$ and $\phi_0$ are varied.
We use the
following flat priors on the 3 free parameters. Firstly, the
orientation $\phi_0$ could in principle be constrained independently
by the shape of the luminous host object. However, there might be a
systematic difference between the host orientation and the dark matter
halo orientation. This misalignment appears to be small for elliptical
galaxies \citep{1998ApJ...509..561K}. We assume the polar angle of the
orientation of the halo is distributed in a range smaller than $\pi/4$
with respect to the known orientation of the host. Secondly, we
restrict the ellipticity to the range between 0.0 and 0.4. Ellipticity
can be constrained using the morphologies of the lensing hosts.
Finally, cluster studies using complementary approaches
from kinematics and X-ray observations should allow us to
independently constrain the halo mass, or $\theta_{\rm E}$. We allow about
$30\%$ uncertainty around the input value and restrict
$\theta_{\rm E}$ to the range $[7'', 13'']$ in our analysis.

\begin{figure}
  \centerline{\scalebox{1.0}
    {\includegraphics[width=6.5cm,height=6.cm]{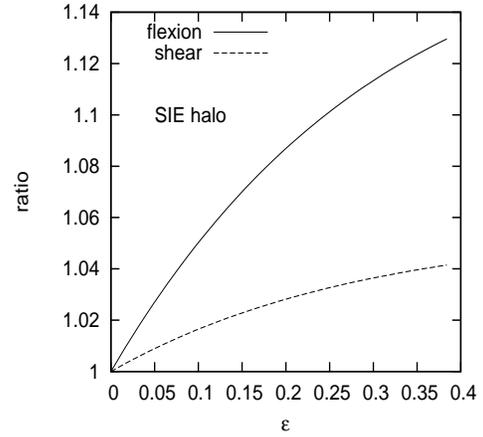}}}
  \caption{Amplitude of shear (dashed) and flexion (solid) at a typical
    location ($\theta_1',\theta_2'$)=($10'',30''$) as function of the
    SIE halo ellipticity, normalized by the amplitude in the singular
    isothermal spherical model.}
  \label{fig:ratiosie}
\end{figure}

Figure~\ref{fig:shearonly} shows the resulting credible contours for the
SIE halo parameters. Both shear and flexion amplitudes increase with halo mass
and also increase, on average over source locations, with halo ellipticity, as
shown in  Fig.~\ref{fig:ratiosie}.
Accordingly, the likelihood analysis produces an anti-correlated contour,
as shown in Fig.~\ref{fig:shearonly} (left panel).
Flexion is more sensitive to a change in the halo ellipticity than shear,
producing tighter constraints. For the SIE model the introduction of
flexion data produces a gain of a factor of 2 on the
 marginalized errors of ellipticity and mass (see Tab.\ref{table}).

There is also a large improvement on the uncertainty of the halo orientation
when using flexion data, as seen in the right panel of Fig.~\ref{fig:shearonly}.
To check the robustness of the result against sample variance we made
various realizations of the noise and location of the background galaxies,
having obtained similar results. We also made tests using a higher number
of background galaxies, obtaining correspondingly tighter constraints.

\begin{figure}
  \centerline{\scalebox{1.0}
    {\includegraphics[width=6.5cm,height=6.cm]{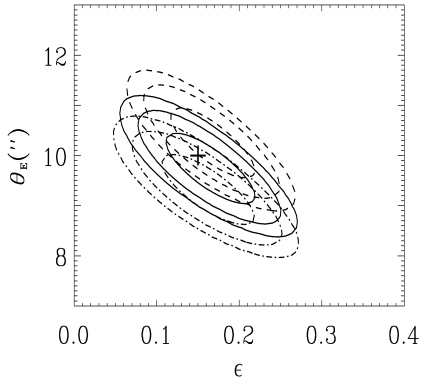}}}
  \caption{SIE $68\%$, $95\%$ and $99\%$ credible regions using
    shear and flexion data, marginalized over redshift uncertainty.  The
    result of Fig.~\ref{fig:shearonly} is shown for comparison (solid
    contours). The cross shows the fiducial input model.}
  \label{fig:zerror}
\end{figure}

These results assume perfectly known redshifts for the source
galaxies.  We introduce now a redshift uncertainty of $2\%$ on average
to simulate photo-z errors, which increase with redshift
\citep{2000A&A...363..476B,2010A&A...523A..31H}.  The marginalized
contours for SIE parameters are shown in Fig.~\ref{fig:zerror}. The
impact of the redshift uncertainty is significant, due to the strong
degeneracy between $\theta_{\rm E}$ and redshift, increasing the
$1\sigma$ error on the $\theta_{\rm E}$ estimate by roughly a factor
of 2.  Furthermore, the redshift uncertainty produces a selection bias
in realizations with excess of low-redshift sources, introducing a
bias in the $\theta_{\rm E}$ estimate. Due to the anti-correlation
found between $\theta_{\rm E}$ and $\epsilon$ in the SIE model, such
selection bias will also affect the ellipticity estimate. On the
other hand, we found no significant effect on the orientation.

\begin{center}
\begin{table*}
\begin{tabular}{|c|c|c|c|c|}
\hline \hline
 &\multicolumn{2}{|c|}{SIE} & \multicolumn{2}{|c|}{eNFW}\\ \hline
  &Shear &Shear + Flexion & Shear & Shear + Flexion\\ \hline
  $\epsilon$  & $0.115\pm 0.061$ &  $0.164\pm 0.035$ &  $0.242 \pm 0.101$& $0.174\pm 0.060$\\ \hline
$\theta_{\rm E} ('')$ &  $10.38 \pm 0.78$ & $9.75 \pm 0.46$ & &\\ \hline
$\kappa_s$ & & & $0.256\pm 0.030$& $0.242\pm 0.022$ \\ \hline
$\delta M$ &  $15\%$ &  $9\%$ &$32\%$ & $26\% $\\ \hline
\hline
\end{tabular}
\caption{\label{table} $1\sigma$ errors estimated for the parameters
  of both models using shear and shear+flexion data. For comparison
  between SIE and eNFW the uncertainties on lens strength parameters
  were converted into mass uncertainties in the last row. The mass of
  SIE model stands for the mass within the Einstein radius
  ($\theta_{\rm E}$), and the mass of eNFW model stands for
  $M_{200}$. }
\end{table*}
\end{center}

\subsection{eNFW model}
In order to test the universality of the result we perform our analysis
on a more realistic model of halo density profile,
adopting an eNFW model for the
foreground lens halo. The NFW profile
\citep{1996ApJ...462..563N,1997ApJ...490..493N} is widely used to
model the halo of galaxy clusters. The dimensionless surface mass
density of a spherical NFW halo is written as
\citep{1996A&A...313..697B,2006MNRAS.365..414B}
\be
\kappa(x) = 2\kappa_s {f(x) \over x^2-1},
\label{kappaenfw}
\ee
where $x$ is the dimensionless radius, the radius $r$ normalized by
the scaling radius $r_s$ ($x\equiv r/r_s$), and the function $f(x)$ is given by
\be
f(x)=
\begin{cases}
  1 - \dfrac{2}{\sqrt{x^2 - 1}}{\rm arctan}\sqrt{\dfrac{x-1}{x+1}} \;\; (x>1);\\
  \\0 \quad\quad\quad\quad\quad\quad\quad\quad\quad\quad\quad \;\; (x=1); \\ \\
  1 - \dfrac{2}{\sqrt{1-x^2}}{\rm arctanh}\sqrt{\dfrac{1-x}{1+x}} \;\; (x<1). \\
\end{cases}
\ee
The physical properties of the halo are contained in the parameter
$\kappa_s=\rho_{\rm
  crit}r_s\Delta_c/\Sigma_{\rm cr}$, where $\Delta_c$ is the
dimensionless characteristic density. The halo mass is defined as
$M_{200}=4/3\rho_{\rm crit}\pi r_{200}^3$, where $r_{200}=r_s\,c$ and $c$ is
the concentration parameter (see the appendix in Navarro et~al. 1997).

For an elliptical halo the dimensionless radius at a point
$(\theta_1,\theta_2)$, defined along the axes of the halo, becomes
\be
x=\frac{\rho}{1-\epsilon}\,\frac{D_{\rm d}}{r_s}.
\ee
The halo ellipticity is defined by
$\epsilon=(\theta_a-\theta_b)/(\theta_a+\theta_b)$ (or equivalently by
the axial ratio $f=\theta_b/\theta_a$), where $\theta_a$, $\theta_b$
are the major and minor axes.

The lensing properties of the eNFW halo can be
calculated numerically given an arbitrary normalized halo convergence
\citep{2001astro.ph..2341K,2009MNRAS.400.1132H} as follows.

The shear and flexion are second and third order derivatives of the lensing potential $\psi$:
\bea
\gamma &=& {1\over 2} (\psi_{11} - \psi_{22}) + {\rm i} \psi_{12};\\
{\cal F} &=& {1\over 2} \eck{\psi_{111} + \psi_{122} + {\rm i} (\psi_{112}
  + \psi_{222}) };\\
{\cal G} &=& {1\over 2} \eck{\psi_{111} - 3\psi_{122} + {\rm i} (3\psi_{112} -
\psi_{222})},
\eea
where subscripts denote partial differentiation. The second and third
order derivatives of the lensing potential, for an eNFW dark matter halo
 at a position $(x,y)$ on the image plane, are given by
\bea
\psi_{11} &=& 2fx^2K_0 + f J_0;\\
\psi_{22} &=& 2fy^2K_0 + f J_1;\\
\psi_{12} &=& 2fxyK_1;\\
\psi_{111} &=& 6fxK_0 + 4fx^3 L_0 ;\\
\psi_{222} &=& 6fyK_2 + 4fy^3 L_3 ;\\
\psi_{112} &=& 2fyK_1 + 4fx^2 yL_1 ;\\
\psi_{122} &=& 2fxK_1 + 4fy^2 xL_2 .
\eea
Here
\bea
J_n(x,y) &=& \int_0^1 \frac{\kappa (\xi (u)^2) \d u}{\eck{1-(1-f^2)u}^{n+1/2}},\\
K_n (x, y) &=& \int_0^1
\frac{u\,\kappa'\,(\xi (u)^2 )\,du} {[1 - (1 - f^2 )u]^{n+1/2}},\\
L_n (x, y) &=& \int_0^1
\frac{u^2\,\kappa''\,(\xi(u)^2) du}{[1-(1-f^2)u]^{n+1/2}},
\eea
are one-dimensional integrals, where $f$ is the axis ratio of the lens
and $\kappa'$ and $\kappa''$ are the first and second order
derivatives of the convergence, e.g. $\kappa'(\xi^2) = \d
\kappa(\xi^2) / \d(\xi^2)$.  The convergence $\kappa$ is written as a
function of the ellipse coordinate $\xi(u)$ given by
\be
\xi(u)^2 = u \rund{x^2 + {y^2 \over 1- (1-f^2) u}}.
\ee
Notice here we need to use $x^2$ instead of $x$ as the independent variable
in $\kappa(x)$.

\begin{figure*}
\centerline{\scalebox{1.0}
  {\includegraphics[width=7.5cm,height=7.cm]{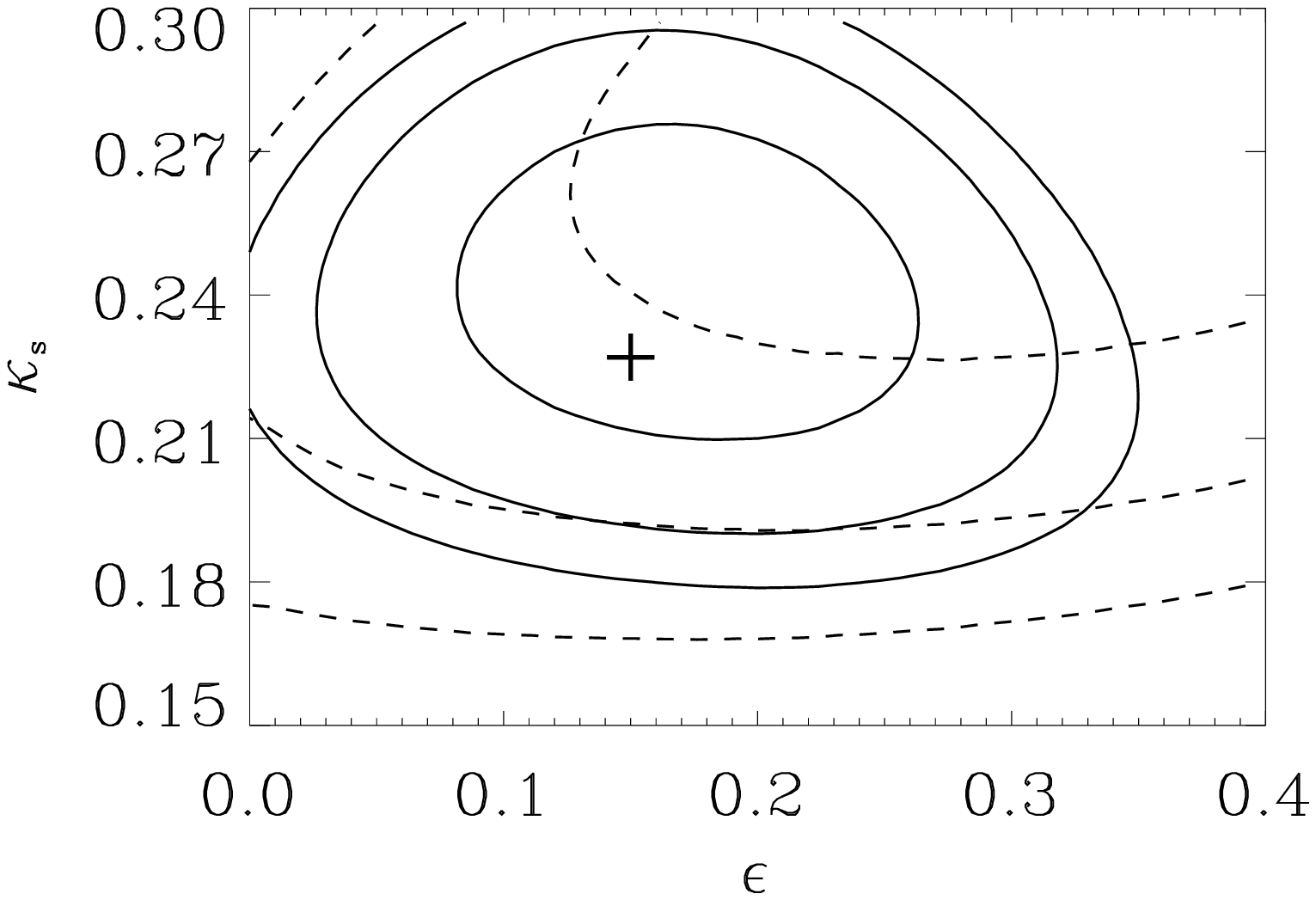}
    \includegraphics[width=7.5cm,height=7.cm]{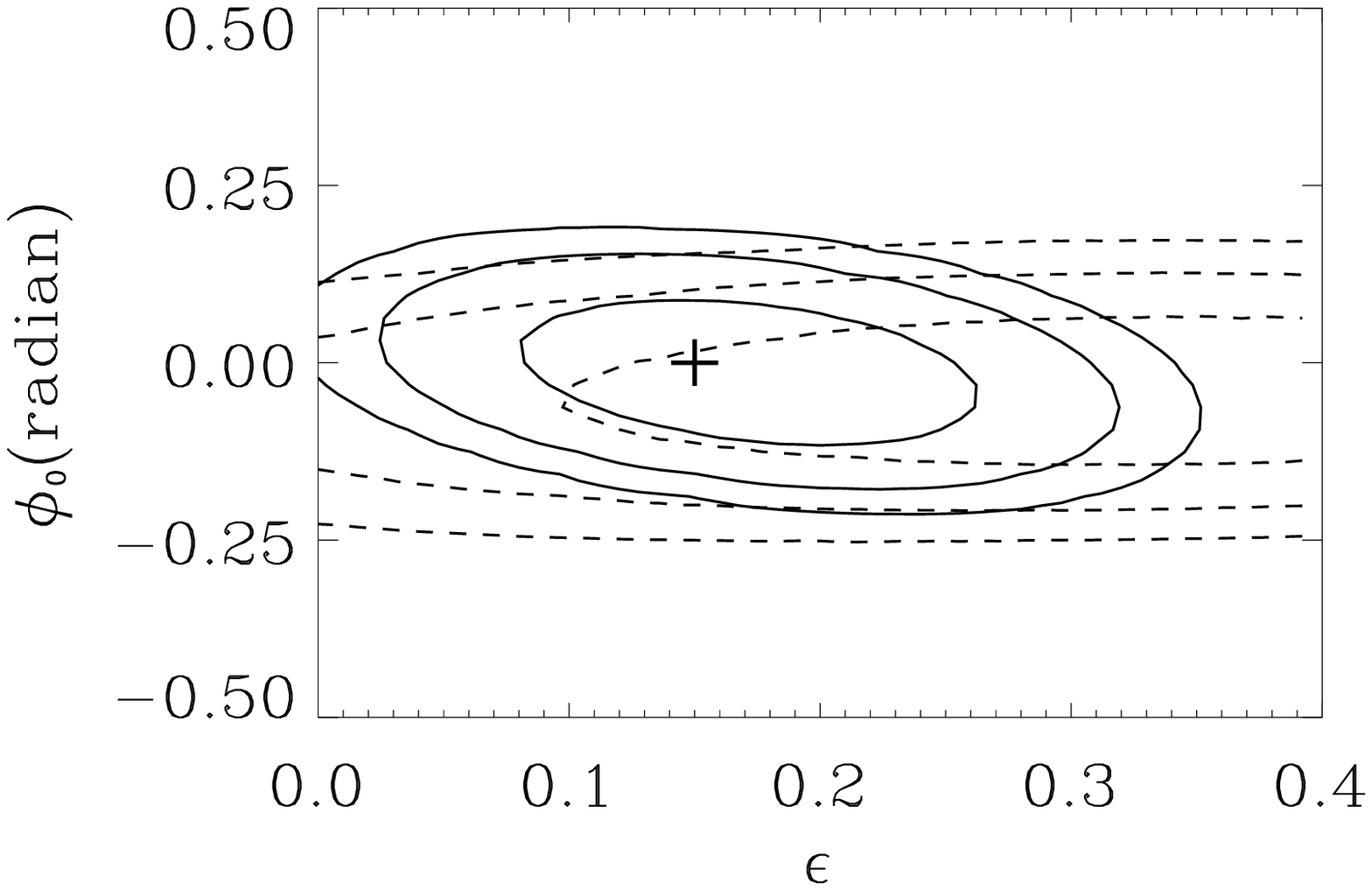}}}
  \caption{Marginalized eNFW $68\%$, $95\%$ and $99\%$ credible regions using
    shear data only (dashed) and combined shear + flexion data (solid).
 Left panel: $(\theta_{\rm E}$-$\epsilon$) plane.
 Right panel: $(\phi_0$-$\epsilon)$ plane. The cross shows the fiducial input model. }
  \label{fig:enfw}
\end{figure*}
\subsection{eNFW results}

In the eNFW model, we consider 3 free parameters:
$(\kappa_s,\epsilon,\phi)$. For our fiducial halo we use a halo mass of
$M_{200} = 1.8\times 10^{14}M_{\odot}$, a concentration
parameter of $c=7.2$, and place the halo at redshift $z_{\rm d}=0.6$. This
implies $\kappa_s=0.227$. For the ellipticity we choose
$\epsilon=0.15$ and for position angle $\phi_0=0$.

The theoretical predictions are calculated numerically and the three
parameters $\kappa_s$, $\epsilon$ and $\phi_0$ are varied.
Similar to the SIE analysis, we place 80 background galaxies
 at random positions in the same field, and randomize their
shear and flexion values. The same filter to the data is employed to
discard the unmeasurable data in the analysis.
We perform a likelihood analysis restricting $\phi_0$ to the range
$[-\pi/4,\pi/4]$, using the range $[0.15, 0.3]$ for $\kappa_s$
and $[0,0.4]$ for the ellipticity $\epsilon$.

Figure~\ref{fig:enfw} shows the resulting credible intervals for the
eNFW halo parameters. Overall, the constraints are looser than in the
SIE model, which has a higher signal-to-noise ratio.
The effective number of flexed background images available for the analysis
 after discarding is lower than that in the SIE analysis by about $15\%$.

The constraints from shear information alone, using priors similar to the
 ones used in the SIE analysis,
are dominated by the priors. Hence, shear data do not add much information to
observations of the luminous host and complementary probes of mass
(e.g. X-ray).

\begin{figure}
  \includegraphics[width=6.5cm,height=6.cm]{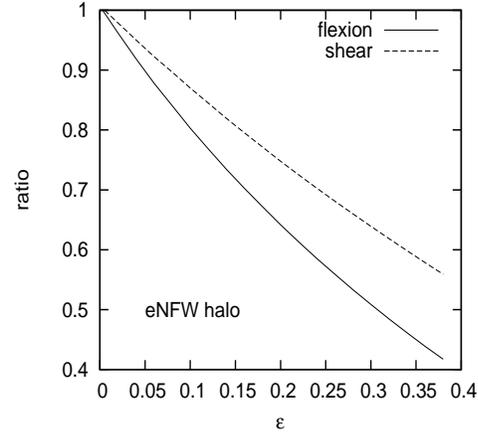}
  \caption{Amplitude of shear (dashed) and flexion (solid) at a
    typical location $(\theta_1',\theta_2')=(10'',30'')$ as a function
    of the eNFW halo ellipticity, normalized by the amplitude in the
    spherical NFW model.}
  \label{fig:ratio}
\end{figure}

Ellipticity and lens strength are now positively correlated in the shear
signal due to the decreasing shear amplitude with ellipticity, as shown in
Fig.~\ref{fig:ratio}. The addition of flexion data decreases the correlation.
Once again, flexion is more sensitive to the change in the halo ellipticity than
shear. Indeed, the tangential shear
mainly depends on the lens strength ($\theta_{\rm E}$ or
$\kappa_s$) and depends weakly on the halo ellipticity, while the cross
shear component is independent of ellipticity. On the other hand, both
flexion components depend on the halo ellipticity and orientation
$\phi_0$. This allows for a tighter constraint when including flexion
information. In Table~\ref{table} the shear and flexion marginalized constraints
 on the ellipticity and strength parameters are roughly a factor of
1.5 tighter than the corresponding shear constraints.
Notice that the effective gain of using
flexion is larger than this factor. Here the shear constraint, contrary to
the combined one, is dominated by the prior; with a wider prior range
the shear constraint would be looser.

We also stress that the eNFW model constraints are not marginalized on all
halo parameters, since the scaling radius $r_s$ is kept fixed in the analysis.
Finally, we checked the results against sample variance making various
realizations of noise and location of the background galaxies,
similar results are found.

\section{Conclusions}

In this paper, we study the potential of weak lensing flexion in the
study of galaxy cluster haloes. We use mock data including shear, $\cal
F$ flexion, $\cal G$ flexion and redshift information.  We find that
the inclusion of flexion significantly improves the estimate of
foreground haloes parameters, although the details are model-dependent.
In particular, in the case of a SIE halo, the presence of a
mass-ellipticity anti-correlation implies that analyses where the halo
is incorrectly assumed to be spherical will overestimate the halo
mass.  On the other hand, we did not find significant correlation
between the halo mass and ellipticity in the eNFW model.

The noise in the mock data is determined by the dispersion of the
intrinsic shear and flexion distributions, and by the density of
background galaxies.  After applying stringent cuts in the data, we
are left with a galaxy density of roughly 10 arcmin$^{-2}$.  Our
approach assumes that the flexion estimators are linear in the flexion
observables and the point spread function can be removed without
producing a bias. In reality, the noise of flexion estimators can be
complicated and non-Gaussian. An accurate study of flexion noise is
important to evaluate the estimated error.

The analysis considers a single cluster
halo. This approach is not possible if the number density of
background galaxies is low. In that case, stacking of several halo
fields can be used to increase the number of background images and
thus the signal-to-noise. That approach requires the alignment of the
major axis of various foreground galaxies and selecting haloes
with similar properties, for example similar shapes of their central
galaxies. Such stacking analysis can constrain the halo shapes more
tightly, and as a function of other halo properties, e.g. mass.

Our results emphasize that a combined weak lensing analysis will
be a useful technique for precise measurements of the properties of
galaxy or cluster haloes from future weak lensing surveys, such as
EUCLID.

\section*{Acknowledgments}

We thank the referee David Goldberg for useful comments on the manuscript.
We also thank Charles Keeton for help with the numerical NFW method.
XE is supported by the Young Researcher Grant of the National
Astronomical Observatories of China. XE and SM thank the Chinese Academy of
Sciences for financial support. IT is funded by FCT
and acknowledges support from the European Programme FP7-PEOPLE-2010-RG-268312.

\bibliographystyle{mn2e}
\bibliography{../../../bib/refbooks,../../../bib/lens,../../../bib/flexion,../../../bib/refcos,../../../bib/shape}

\end{document}